\documentclass[11pt]{article}
\usepackage{graphics}
\usepackage{enumerate}
\usepackage{amssymb}
\usepackage{amscd}
\usepackage[intlimits]{amsmath}
\usepackage{graphicx}
\usepackage{exscale}

\usepackage{makeidx,xspace}

\usepackage[usenames]{color}
\usepackage{url}

\usepackage{amsmath,amsfonts,amssymb,amsbsy}
\usepackage{natbib}
\usepackage[cp1251]{inputenc}
\usepackage[english]{babel}

\long\def\symbolfootnote[#1]#2{\begingroup%
\def\thefootnote{\fnsymbol{footnote}}\footnote[#1]{#2}\endgroup}

\usepackage{makeidx,xspace}

\usepackage[usenames]{color}
\usepackage{url}

\newfont\Bbbb{msbm9 scaled 1200}
\newfont\smallBbb{msbm8 scaled 1000}
\begin{document}

\centerline{\large\bf
A closed-form solution of the three-dimensional contact problem}
\centerline{\large\bf for biphasic cartilage layers}
\medskip
\centerline{I.~I.~Argatov,
G.~S.~Mishuris\symbolfootnote[1]{Corresponding author. E-mail: {\tt
ggm@aber.ac.uk}}}
\medskip
\centerline{\it Institute of Mathematics and Physics, Aberystwyth University,}
\centerline{\it Ceredigion SY23 3BZ, Wales, UK}
\bigskip
\medskip

{\bf Abstract:}
A three-dimensional unilateral contact problem for articular cartilage layers is considered in the framework of the biphasic cartilage model. The articular cartilages bonded to subchondral bones are modeled as biphasic materials consisting of a solid phase and a fluid phase. It is assumed that the subchondral bones are rigid and shaped like elliptic paraboloids. The obtained analytical solution is valid over long time periods and can be used for increasing loading conditions.

\medskip


\bigskip

\setcounter{equation}{0}

\section*{Introduction}

Biomechanical contact problems involving transmission of forces across biological joints are of considerable practical importance in surgery. Many solutions to the axisymmetric problem of contact interaction of articular cartilage surfaces in joints are available. \citet{Ateshian1994} obtained an asymptotic solution for the contact problem of two identical biphasic cartilage
layers attached to two rigid impermeable spherical bones of equal radii modeled as elliptic paraboloids.
\citet{Wu1996} extended this solution to a more general model by combining the assumption of the kinetic relationship from classical contact mechanics \citep{Johnson} with the joint contact model for the contact of two biphasic cartilage \citep{Ateshian1994}. An improved solution for the contact of two biphasic cartilage layers which can be used for dynamic loading was obtained by \citet{Wu1997}. These solutions have been widely used as theoretical background in modeling the articular contact mechanics. Recently, \citet{MishurisArgatov2009,ArgatovMishuris2010} extended the analysis of \citet{Wu1996} by formulating the refined contact condition which takes into account the tangential displacements at the contact region.

When studying contact problems for real joint geometries, a numerical analysis, such as the finite element method, is necessary \citep{Han_et_al2005}, since exact analytical solutions can be only be obtained for two-dimensional \citep{AteshianWang(1995}, or axisymmetric and simple geometries \citep{Eberhardt_et_al1990,Eberhardt_et_al1991,Li_et_al1997}. In this study, the axisymmetric model of articular contact mechanics developed by \citet{Ateshian1994,Wu1996} is generalized for the three-dimensional case. The method developed by \citet{Arg2004ref} is used to obtain general relationships between the integral characteristics of the contact problem. The exact closed-form solution of the contact problem for biphasic cartilage layers attached to rigid bones shaped like elliptic paraboloids is obtained.

\section{Formulation of the contact problem}

We consider a frictionless contact between two thin linear biphasic cartilage layers firmly attached to rigid bones shaped like elliptic paraboloids. Introducing the Cartesian coordinate system $(x_1,x_2,x_3)$, we write the equations of the cartilage surfaces (before loading) in the form $x_3=(-1)^n\Phi_n(x_1,x_2)$ ($n=1,2$). We assume that the two cartilage-bone systems occupy convex domains $x_3\leq -\Phi_1(x_1,x_2)$ and $x_3\geq \Phi_2(x_1,x_2)$ whereas in the undeformed state they are in contact with the plane $x_3=0$ at a single point chosen as the coordinate origin. In the particular case of bones shaped like elliptic paraboloids, we have
\begin{equation}
\Phi_n(x_1,x_2)=\frac{x_1^2}{2R_1^{(n)}}+\frac{x_2^2}{2R_2^{(n)}}\quad (n=1,2),
\label{3Dc(1.7)}
\end{equation}
where $R_1^{(n)}$ and $R_2^{(n)}$ are the curvature radii of the $n$-th bone surface at its apex.

We denote the vertical approach of the bones by $-\delta_0(t)$. Then, the linearized unilateral contact condition that the boundary points of the cartilage layers do not penetrate one into another can be written as follows:
\begin{equation}
\delta_0(t)-w_1(x_1,x_2,t)-w_2(x_1,x_2,t)\leq \Phi_1(x_1,x_2)+\Phi_2(x_1,x_2).
\label{3Dc(1.1)}
\end{equation}

An asymptotic solution obtained by \citet{Ateshian1994} for the vertical displacement of the boundary points of a biphasic cartilage layer, $w_n(x_1,x_2,t)$, in the axisymmetric problem of acting contact pressure on its surface can be generalized for the three-dimensional case as follows:
\begin{equation}
w_n(x_1,x_2,t) = \frac{h_n^3}{3\mu_{sn}}
\biggl\{\Delta P(x_1,x_2,t)+\frac{3\mu_{sn}k_n}{h_n^2}\int\limits_0^{t}
\Delta P(x_1,x_2,\tau)\,d\tau\biggr\}.
\label{3Dc(1.3)}
\end{equation}
Here, $\mu_{sn}$ is the shear modulus of the solid phase of the cartilage tissue ($n=1,2$), $h_1$ and $h_2$ are the thicknesses of the cartilage layers, $k_1$ and $k_2$ are the cartilage permeabilities, $P(x_1,x_2,t)$ is the contact pressure, $\Delta=\partial^2/\partial x_1^2+\partial^2/\partial x_2^2$ is the Laplace operator.

The equality in relation (\ref{3Dc(1.1)}) determines the contact region $\omega(t)$. In other words, the following equation holds within the contact area:
\begin{equation}
w_1(x_1,x_2,t)+w_2(x_1,x_2,t)=\delta_0(t)-\Phi(x_1,x_2),\quad (x_1,x_2)\in \omega(t).
\label{3Dc(1.5)}
\end{equation}
Here we introduced the notation
\begin{equation}
\Phi(x_1,x_2)=\Phi_1(x_1,x_2)+\Phi_2(x_1,x_2).
\label{3Dc(1.1Ph)}
\end{equation}

Note that in the case (\ref{3Dc(1.7)}), Eq.~(\ref{3Dc(1.1Ph)}) takes the form
\begin{equation}
\Phi(x_1,x_2)=\frac{x_1^2}{2R_1}+\frac{x_2^2}{2R_2},
\label{3Dc(1.1Ph2)}
\end{equation}
where the parameters $R_1$ and $R_2$ are determined by the formulas
$$
\frac{1}{R_1}=\frac{1}{R_1^{(1)}}+\frac{1}{R_1^{(2)}},\quad
\frac{1}{R_2}=\frac{1}{R_2^{(1)}}+\frac{1}{R_2^{(2)}}.
$$

Substituting the expressions for the displacements $w_1(x_1,x_2,t)$ and $w_2(x_1,x_2,t)$ given by formula (\ref{3Dc(1.3)}) into Eq.~(\ref{3Dc(1.5)}), we obtain the contact condition in the following form (we assume that $(x_1,x_2)\in\omega(t)$):
\begin{equation}
\Delta P(x_1,x_2,t)+\chi\int\limits_0^t \Delta P(x_1,x_2,\tau)\,d\tau =
m\bigl(\Phi(x_1,x_2)-\delta_0(t)\bigr).
\label{3Dc(1.9)}
\end{equation}
Here we introduced the notation
\begin{equation}
\chi=\frac{3\mu_{s1} k_1}{h^2_1}+\frac{3\mu_{s2} k_2}{h^2_2},\quad
m=\biggl(\frac{h_1^3}{3\mu_{s1}}+\frac{h_2^3}{3\mu_{s2}}\biggr)^{-1}.
\label{3Dc(1.10)}
\end{equation}

Eq.~(\ref{3Dc(1.9)}) will be used to find the contact pressure density $P(x_1,x_2,t)$. The contour $\Gamma(t)$ of the contact area $\omega(t)$ is determined from the condition that the contact pressure is positive and vanishes at the contour of the contact area:
\begin{equation}
P(x_1,x_2,t)\geq 0, \quad (x_1,x_2)\in\omega(t);\quad P(x_1,x_2,t)=0, \quad (x_1,x_2)\in\Gamma(t).
\label{3Dc(1.11)}
\end{equation}
Moreover, in the case of contact problem for a biphasic cartilage layer, in which the contact pressure is carried primarily by the fluid phase, it is additionally assumed a smooth transition of the surface normal stresses from the contact region $(x_1,x_2)\in\omega(t)$ to the outside region $(x_1,x_2)\not\in\omega(t)$ \citep{Ateshian1994}.
Thus, we impose the following boundary condition:
\begin{equation}
\frac{\partial P}{\partial n}(x_1,x_2,t)=0, \quad (x_1,x_2)\in\Gamma(t).
\label{3Dc(1.12)}
\end{equation}
Here, $\partial/\partial n$ is the normal derivative directed outward from $\omega(t)$.

We assume that the density $P(x_1,x_2,t)$ is defined on the entire plane such that
\begin{equation}
P(x_1,x_2,t)= 0, \quad (x_1,x_2)\not\in\omega(t).
\label{3Dc(1.11b)}
\end{equation}

Finally, from the physical point of view, the contact pressure under a blunt punch with a smooth surface should satisfy the regularity condition, i.\,e., in the case (\ref{3Dc(1.1Ph2)}), the function $P(x_1,x_2,t)$ is assumed to be analytical in the domain $\omega(t)$.

The equilibrium equation for the whole system is
\begin{equation}
\iint\limits_{\omega(t)} P(x_1,x_2,t)\,dx_1 dx_2=F(t),
\label{3Dc(1.14)}
\end{equation}
where $F(t)$ denotes the external load.

For non-decreasing loads when $dF(t)/dt\geq 0$, the contact zone should increase. Thus, we assume that the following monotonicity condition holds:
\begin{equation}
\omega(t_1)\subset \omega(t_2),\quad t_1\leq t_2.
\label{3Dc(1.15)}
\end{equation}

The aim of this study is to derive an asymptotic solution for the three-dimensional contact problem for biphasic cartilage layers formulated by Eq.~(\ref{3Dc(1.9)}) under the monotonicity condition (\ref{3Dc(1.15)}). Notice that in the axisymmetric case the contact problem under consideration coincides with that studied in detail by \citet{Ateshian1994}, \citet{Wu1997}.

\section{Equation for the displacement parameter}

Integrating Eq.~(\ref{3Dc(1.9)}) over the contact domain $\omega(t)$, we get
\begin{equation}
\iint\limits_{\omega(t)}\Delta P({\bf y},t)\,d{\bf y}
+\chi \iint\limits_{\omega(t)}\int\limits_0^t
\Delta P({\bf y},\tau)\,d{\bf y}d\tau =
m\iint\limits_{\omega(t)}
\bigl(\Phi({\bf y})-\delta_0(t)\bigr)\,d{\bf y}.
\label{3Dc(2.1)}
\end{equation}
Here we used the notation ${\bf y}=(y_1,y_2)$ and $d{\bf y}=d y_1 d y_2$.

In view of (\ref{3Dc(1.11b)}) and (\ref{3Dc(1.15)}), we have $\omega(\tau)\subset \omega(t)$ and
$P({\bf y},\tau)\equiv 0$ for ${\bf y}\not\in\omega(\tau)$. Therefore, the second integral on the left-hand side of (\ref{3Dc(2.1)}) takes the form
\begin{equation}
\iint\limits_{\omega(t)}\int\limits_0^t
\Delta P({\bf y},\tau)\,d{\bf y}d\tau =\int\limits_0^t \iint\limits_{\omega(\tau)}
\Delta P({\bf y},\tau)\,d{\bf y}d\tau .
\label{3Dc(2.1a)}
\end{equation}
Note that the density $P(x_1,x_2,t)$ is a smooth function of the variables $x_1$ and $x_2$ on the entire plane.

Using the second Green's formula
\begin{equation}
\iint\limits_{\omega(t)}
\bigl(u({\bf y})\Delta v({\bf y})-v({\bf y})\Delta u({\bf y})\bigr)\,d{\bf y} =
\int\limits_{\Gamma(t)}\biggl(
u({\bf y})\frac{\partial v}{\partial n}({\bf y})-
v({\bf y})\frac{\partial u}{\partial n}({\bf y})
\biggr)ds,
\label{3Dc(2.1G)}
\end{equation}
where $ds$ is the element of the arc length, we obtain
\begin{equation}
\iint\limits_{\omega(\tau)} \Delta P({\bf y},\tau)\,d{\bf y}=
\int\limits_{\Gamma(\tau)}\frac{\partial P}{\partial n}({\bf y},\tau)\,ds.
\label{3Dc(2.1b)}
\end{equation}

Thus, taking into account formulas (\ref{3Dc(2.1a)}) and (\ref{3Dc(2.1b)}), we rewrite Eq.~(\ref{3Dc(2.1)}) as follows:
\begin{equation}
\int\limits_{\Gamma(t)}\frac{\partial P}{\partial n}({\bf y},t)\,ds
+\chi \int\limits_0^t
\int\limits_{\Gamma(\tau)}\frac{\partial P}{\partial n}({\bf y},\tau)\,ds d\tau=
m\iint\limits_{\omega(t)}\Phi({\bf y})\,d{\bf y}
-m A(t)\delta_0(t).
\label{3Dc(2.2)}
\end{equation}
Here, $A(t)$ is the area of $\omega(t)$ given by the integral
\begin{equation}
A(t)=\iint\limits_{\omega(t)}d{\bf y}.
\label{3Dc(2.2A)}
\end{equation}

Finally, in view of the boundary condition (\ref{3Dc(1.12)}), from Eq.~(\ref{3Dc(2.2)}) it follows that
\begin{equation}
\delta_0(t)=\frac{1}{A(t)}\iint\limits_{\omega(t)}\Phi({\bf y})\,d{\bf y}.
\label{3Dc(2.3)}
\end{equation}
Eq.~(\ref{3Dc(2.3)}) connects the unknown displacement parameter $\delta_0(t)$ with the integral characteristic of the contact domain $\omega(t)$. In the case of the axisymmetric problem it coincides with the results obtained by \citet{Ateshian1994,Wu1997}.

\section{Equation for the integral characteristics the contact domain}

Substituting the functions $u(x_1,x_2)=P(x_1,x_2,t)$ and $v(x_1,x_2)=(1/4)(x_1^2+x_2^2)$ into Green's formula (\ref{3Dc(2.1G)}) and taking into account the boundary conditions (\ref{3Dc(1.11)}) and (\ref{3Dc(1.12)}), we obtain the relation
\begin{equation}
\frac{1}{4}\iint\limits_{\omega(t)}
\vert{\bf y}\vert^2\Delta P({\bf y},t)\,d{\bf y}=
\iint\limits_{\omega(t)} P({\bf y},t)\,d{\bf y}.
\label{3Dc(2.4)}
\end{equation}

Using formula (\ref{3Dc(2.4)}), we can evaluate the contact load (\ref{3Dc(1.14)}). Indeed, multiplying the both sides of Eq.~(\ref{3Dc(1.9)}) by $(1/4)(x_1^2+x_2^2)$ and integrating the obtained equation over the contact domain $\omega(t)$, we obtain
\begin{equation}
\iint\limits_{\omega(t)} P({\bf y},t)\,d{\bf y}
+\chi \int\limits_0^t \iint\limits_{\omega(\tau)}
P({\bf y},\tau)\,d{\bf y}d\tau =
\frac{m}{4} \iint\limits_{\omega(t)}\vert{\bf y}\vert^2
\bigl(\Phi({\bf y})-\delta_0(t)\bigr)\, d{\bf y}.
\label{3Dc(2.5)}
\end{equation}

Taking the notation (\ref{3Dc(1.14)}) into account, we rewrite Eq.~(\ref{3Dc(2.5)}) as follows:
\begin{equation}
F(t)+\chi \int\limits_0^t F(\tau)\,d\tau =
\frac{m}{4}
\iint\limits_{\omega(t)}\vert{\bf y}\vert^2\Phi({\bf y})\, d{\bf y}
-\delta_0(t)\frac{m}{4}
\iint\limits_{\omega(t)}\vert{\bf y}\vert^2 d{\bf y}.
\label{3Dc(2.6)}
\end{equation}

Excluding the quantity $\delta_0(t)$ from Eq.~(\ref{3Dc(2.6)}) by means of Eq.~(\ref{3Dc(2.3)}), we derive the following equation:
\begin{equation}
F(t)+\chi \int\limits_0^t F(\tau)\,d\tau =
\frac{m}{4} \iint\limits_{\omega(t)}\biggl(
\vert{\bf y}\vert^2-\frac{J_0(t)}{A(t)}\biggr)\Phi({\bf y})\, d{\bf y}.
\label{3Dc(2.7)}
\end{equation}
Here, $J_0(t)$ is the polar moment of inertia of $\omega(t)$ given by the integral
\begin{equation}
J_0(t)=\iint\limits_{\omega(t)}\vert{\bf y}\vert^2 d{\bf y}.
\label{3Dc(2.2J0)}
\end{equation}

Eq.~(\ref{3Dc(2.7)}) connects the integral characteristic of the unknown contact domain $\omega(t)$ and the known contact load $F(t)$. In the case of the axisymmetric problem it coincides with the results obtained by \citet{Wu1997}.

\section{Contact domain}

Let us rewrite Eq.~(\ref{3Dc(1.9)}) in the form
\begin{equation}
\Delta p(x_1,x_2,t)=m\bigl(\Phi(x_1,x_2)-\delta_0(t)\bigr),
\label{3Dc(3.1)}
\end{equation}
where we introduced the notation
\begin{equation}
p(x_1,x_2,t)=P(x_1,x_2,t)+\chi\int\limits_0^t P(x_1,x_2,\tau)\,d\tau.
\label{3Dc(3.2)}
\end{equation}

In view of the boundary conditions (\ref{3Dc(1.11)}) and (\ref{3Dc(1.12)}), the function $p(x_1,x_2,t)$ must satisfy the following boundary conditions:
\begin{equation}
p(x_1,x_2,t)=0, \quad (x_1,x_2)\in\Gamma(t),
\label{3Dc(3.3)}
\end{equation}
\begin{equation}
\frac{\partial p}{\partial n}(x_1,x_2,t)=0, \quad (x_1,x_2)\in\Gamma(t).
\label{3Dc(3.4)}
\end{equation}

In the case (\ref{3Dc(1.1Ph2)}), we put
\begin{equation}
p(x_1,x_2,t)=p_0(t)\biggl(1-\frac{x_1^2}{a^2(t)}-\frac{x_2^2}{b^2(t)}
\biggr)^2. \label{3Dc(3.5)}
\end{equation}

The representation (\ref{3Dc(3.5)}) assumes that the contour $\Gamma(t)$ is an ellipse with the semi-axises $a(t)$ and $b(t)$. It is not hard to check that the function (\ref{3Dc(3.5)}) satisfies the boundary conditions (\ref{3Dc(3.3)}) and (\ref{3Dc(3.4)}) exactly.

Substituting (\ref{3Dc(3.5)}) into Eq.~(\ref{3Dc(3.1)}), we obtain after some algebra the following system of algebraic equations:
\begin{eqnarray}
\delta_0(t) &=& \frac{4p_0(t)}{m}\biggl(\frac{1}{a^2(t)}+\frac{1}{b^2(t)}\biggr),
\label{3Dc(3.61)}
\\
\frac{1}{2R_1} &=& \frac{4p_0(t)}{m a^2(t)}\biggl(\frac{3}{a^2(t)}+\frac{1}{b^2(t)}\biggr),
\label{3Dc(3.62)}
\\
\frac{1}{2R_2} &=& \frac{4p_0(t)}{m b^2(t)}\biggl(\frac{1}{a^2(t)}+\frac{3}{b^2(t)}\biggr).
\label{3Dc(3.63)}
\end{eqnarray}

The form of the ellipse $\Gamma(t)$ can be characterized by its aspect ratio $s$ defined as follows:
\begin{equation}
s=\frac{b(t)}{a(t)}.
\label{3Dc(3.7)}
\end{equation}

From Eqs.~(\ref{3Dc(3.62)}) and (\ref{3Dc(3.63)}), it immediately follows that
\begin{equation}
\frac{R_2}{R_1}=\frac{s^2(3s^2+1)}{3+s^2}.
\label{3Dc(3.8)}
\end{equation}

Eq.~(\ref{3Dc(3.8)}) can be reduced to a quadratic equation for $s^2$. In this way one can obtain
\begin{equation}
s^2=\sqrt{\biggl(\frac{R_1-R_2}{6R_1}\biggr)^2+\frac{R_2}{R_1}}
-\frac{(R_1-R_2)}{6R_1}.
\label{3Dc(3.9)}
\end{equation}

Further, Eq.~(\ref{3Dc(2.3)}) takes the form
\begin{equation}
\delta_0(t)=\frac{1}{8}\biggl(\frac{1}{R_1}+\frac{s^2}{R_2}\biggr)a^2(t),
\label{3Dc(2.3Ph)}
\end{equation}

Excluding the quantity $\delta_0(t)$ from Eqs.~(\ref{3Dc(3.61)}) and (\ref{3Dc(2.3Ph)}), we obtain
\begin{equation}
p_0(t)=\frac{m}{32}\frac{s^2}{(s^2+1)}\biggl(\frac{1}{R_1}+\frac{s^2}{R_2}\biggr)a^4(t).
\label{3Dc(3.11)}
\end{equation}

Finally, Eq.~(\ref{3Dc(2.7)}) becomes
\begin{equation}
F(t)+\chi \int\limits_0^t F(\tau)\,d\tau =
\frac{m\pi}{384} \biggl(\frac{3s-s^3}{R_1}+\frac{3s^5-s^3}{R_2}\biggr)a^6(t).
\label{3Dc(2.7Ph)}
\end{equation}

Thus, Eq.~(\ref{3Dc(2.7Ph)}) allows to determine the major semi-axis $a(t)$ of the contact domain as a function of time $t$ as follows:
\begin{equation}
a(t)=\biggl[\frac{m\pi}{384}
\biggl(\frac{3s-s^3}{R_1}+\frac{3s^5-s^3}{R_2}\biggr)\biggr]^{-1/6}
\Biggl(F(t)+\chi \int\limits_0^t F(\tau)\,d\tau \Biggr)^{1/6}.
\label{3Dc(2.7Phb)}
\end{equation}

Now, formulas (\ref{3Dc(2.3Ph)}) and (\ref{3Dc(3.11)}) allow to determine the quantities $\delta_0(t)$ and $p_0(t)$, respectively.
Again, in the case of the axisymmetric problem $s=1$, Eq.~(\ref{3Dc(2.7Phb)})  coincides with the corresponding result obtained by \citet{Wu1997}.

\section{Contact pressure}

Let us now introduce the following short hand notation for the operator on the left-hand side of Eq.~(\ref{3Dc(3.2)}):
\begin{equation}
\mathcal{K}y(t)=y(t)+\chi\int\limits_0^t y(\tau)\,d\tau.
\label{3Dc(8.6)}
\end{equation}
The inverse operator to ${\mathcal K}$ denoted by ${\mathcal K}^{-1}$ is defined by the formula
\begin{equation}
{\mathcal K}^{-1}Y(t)=Y(t)-\chi\int\limits_0^t Y(\tau) e^{-\chi (t-\tau)}d\tau.
\label{3Dc(8.7)}
\end{equation}

In view of (\ref{3Dc(8.6)}) and (\ref{3Dc(3.5)}), we obtain the following operator equation for the contact pressure density $P(x_1,x_2,t)$:
\begin{equation}
\mathcal{K}P(x_1,x_2,t)=p_0(t)\biggl(1-\frac{x_1^2}{a^2(t)}-\frac{x_2^2}{b^2(t)}
\biggr)^2, \quad (x_1,x_2)\in\omega(t).
\label{3Dc(8.8)}
\end{equation}

Taking the relation (\ref{3Dc(8.6)}), a solution of Eq.~(\ref{3Dc(8.8)}) can be represented as follows:
\begin{equation}
P(x_1,x_2,t)=\mathcal{K}^{-1}\Biggl\{\biggl(1-\frac{x_1^2}{a^2(t)}-\frac{x_2^2}{b^2(t)}
\biggr)^2p_0(t)\Biggr\},
\label{3Dc(8.11)}
\end{equation}
or in view of the notation (\ref{3Dc(8.7)})
\begin{eqnarray}
P(x_1,x_2,t) & = & \biggl(1-\frac{x_1^2}{a^2(t)}-\frac{x_2^2}{b^2(t)}
\biggr)^2p_0(t) \nonumber \\
{ } & - & \chi\int\limits_0^t
\biggl(1-\frac{x_1^2}{a^2(\tau)}-\frac{x_2^2}{b^2(\tau)}
\biggr)^2
H\biggl(1-\frac{x_1^2}{a^2(\tau)}-\frac{x_2^2}{b^2(\tau)}
\biggr)p_0(\tau) e^{-\chi (t-\tau)}d\tau.
\label{3Dc(8.12)}
\end{eqnarray}
Here, $H(x)$ is the Heaviside step function defined as $H(x)=1$ for $x>0$ and
$H(x)=0$ for $x\leq 0$.

It is clear that if the point $(x_1,x_2)$ belongs to the initial contact zone, i.\,e.,
$$
1-\frac{x_1^2}{a^2(0)}-\frac{x_2^2}{b^2(0)}>0,
$$
then formula (\ref{3Dc(8.12)}) simplifies to
\begin{eqnarray}
P(x_1,x_2,t) & = & \biggl(1-\frac{x_1^2}{a^2(t)}-\frac{x_2^2}{b^2(t)}
\biggr)^2p_0(t) \nonumber \\
{ } & - & \chi\int\limits_0^t
\biggl(1-\frac{x_1^2}{a^2(\tau)}-\frac{x_2^2}{b^2(\tau)}
\biggr)^2 p_0(\tau) e^{-\chi (t-\tau)}d\tau.
\label{3Dc(8.13)}
\end{eqnarray}

If the point $(x_1,x_2)$ lies outside of the initial contact zone, i.\,e.,
$$
1-\frac{x_1^2}{a^2(0)}-\frac{x_2^2}{b^2(0)}<0,
$$
then formula (\ref{3Dc(8.12)}) can be rewritten as
\begin{eqnarray}
P(x_1,x_2,t) & = & \biggl(1-\frac{x_1^2}{a^2(t)}-\frac{x_2^2}{b^2(t)}
\biggr)^2p_0(t) \nonumber \\
{ } & - & \chi\int\limits_{t_*(x_1,x_2)}^t
\biggl(1-\frac{x_1^2}{a^2(\tau)}-\frac{x_2^2}{b^2(\tau)}
\biggr)^2 p_0(\tau) e^{-\chi (t-\tau)}d\tau,
\label{3Dc(8.14)}
\end{eqnarray}
where $t_*(x_1,x_2)$ is the time when the contour of the contact zone first reaches the point $(x_1,x_2)$. The quantity $t_*(x_1,x_2)$ is determined by the equation
$$
a^2(t_*)=x_1^2+\frac{x_2^2}{s^2},
$$
or in accordance with Eq.~(\ref{3Dc(2.3Ph)}) by the following one:
\begin{equation}
F(t_*)+\chi \int\limits_0^{t_*} F(\tau)\,d\tau
=\frac{m\pi}{384}
\biggl(\frac{3s-s^3}{R_1}+\frac{3s^5-s^3}{R_2}\biggr)
\biggl(x_1^2+\frac{x_2^2}{s^2}\biggr)^3.
\label{3Dc(8.15)}
\end{equation}

In the case of a stepwise loading, we have $F(t)=F_0$, and Eq.~(\ref{3Dc(8.15)}) admits the following closed-form solution:
\begin{equation}
t_*(x_1,x_2)=\frac{m\pi}{384\chi F_0}
\biggl(\frac{3s-s^3}{R_1}+\frac{3s^5-s^3}{R_2}\biggr)\biggl[
\biggl(x_1^2+\frac{x_2^2}{s^2}\biggr)^3-a^6(0)\biggr],
\label{3Dc(8.16)}
\end{equation}
where $a(0)$ is the initial value of the major semi-axis of contact domain, and the quantity $a^6(0)$ is given by
\begin{equation}
a^6(0)=\frac{384}{m\pi}
\biggl(\frac{3s-s^3}{R_1}+\frac{3s^5-s^3}{R_2}\biggr)^{-1}F_0.
\label{3Dc(8.16b)}
\end{equation}

Finally, using the Heaviside function and taking into account Eq.~(\ref{3Dc(8.16b)}), we can rewrite Eq.~(\ref{3Dc(8.16)}) in the form
\begin{equation}
t_*(x_1,x_2)=\frac{1}{\chi}
\biggl(
\frac{(x_1^2+s^{-2}x_2^2)^3}{a^6(0)}-1\biggr)
H\bigl(
(x_1^2+s^{-2}x_2^2)^3-a^6(0)\bigr).
\label{3Dc(8.17)}
\end{equation}

Thus, in the case of a stepwise loading, formula (\ref{3Dc(8.14)}), where quantity $t_*(x_1,x_2)$ is determined by Eq.~(\ref{3Dc(8.17)}), represents the sought for solution of Eq.~(\ref{3Dc(1.9)}) in the case of the gap between the contacting surfaces shaped as the elliptic paraboloid (\ref{3Dc(1.1Ph2)}).
Note that in the case of the axisymmetric problem the derived expression for the contact pressure coincides with the result obtained previously by \citet{ArgatovMishuris2010}.

\section*{Conclusion}

The present study results in the exact closed-form solution to the three-dimensional contact problem for biphasic cartilage layers. The general equations (\ref{3Dc(2.3)}) and (\ref{3Dc(2.7)}) as well formulas (\ref{3Dc(3.9)}), (\ref{3Dc(2.7Phb)}), (\ref{3Dc(2.3Ph)}), (\ref{3Dc(3.11)}), (\ref{3Dc(8.12)}) for evaluating the aspect ratio of the elliptic contact domain, its major semi-axis $a(t)$, the displacement parameter $\delta_0(t)$, the auxiliary parameter $p_0(t)$, and the contact pressure $P(x_1,x_2,t)$ in the spacial case (\ref{3Dc(1.1Ph2)}) of contact of elliptic paraboloids constitute the main result of the present study.
The obtained results generalize the solution of \citet{Wu1997} for the elliptic contact of biphasic cartilage layers.

\end{document}